\documentclass[prb,twocolumn,showpacs,showkeys,superscriptaddress,amsmath,amssymb]{revtex4}
\usepackage[dvips]{graphicx}
\usepackage{multirow,slashbox}
\usepackage{ulem}
\DeclareTextSymbol{\degre}{T1}{6}
\DeclareTextSymbol{\degre}{OT1}{23}

\begin{document}

\title{Magnetic and dielectric properties in the langasite-type compounds : A$_3$BFe$_3$D$_2$O$_{14}$ with A=Ba, Sr, Ca, B=Ta, Nb, Sb and D=Ge, Si }

\author{K. Marty}
\affiliation{Institut NEEL\\
CNRS \& UJF, BP 166, 38042 Grenoble Cedex 9, France}

\author{P. Bordet}
\email{pierre.bordet@grenoble.cnrs.fr}
\affiliation{Institut NEEL\\
CNRS \& UJF, BP 166, 38042 Grenoble Cedex 9, France}

\author{V. Simonet}
\affiliation{Institut NEEL\\
CNRS \& UJF, BP 166, 38042 Grenoble Cedex 9, France}

\author{M. Loire}
\affiliation{Institut NEEL\\
CNRS \& UJF, BP 166, 38042 Grenoble Cedex 9, France}

\author{R. Ballou}
\affiliation{Institut NEEL\\
CNRS \& UJF, BP 166, 38042 Grenoble Cedex 9, France}

\author{C. Darie}
\affiliation{Institut NEEL\\
CNRS \& UJF, BP 166, 38042 Grenoble Cedex 9, France}

\author{J. Kljun}
\affiliation{Institut NEEL\\
CNRS \& UJF, BP 166, 38042 Grenoble Cedex 9, France}

\author{P. Bonville}
\affiliation{CEA-Saclay, IRAMIS/Service de Physique de l'Etat Condens\'e, 91191 Gif-sur-Yvette, France}

\author{O. Isnard}
\affiliation{Institut NEEL\\
CNRS \& UJF, BP 166, 38042 Grenoble Cedex 9, France}

\author{P. Lejay}
\affiliation{Institut NEEL\\
CNRS \& UJF, BP 166, 38042 Grenoble Cedex 9, France}

\author{B. Zawilski}
\affiliation{Institut NEEL\\
CNRS \& UJF, BP 166, 38042 Grenoble Cedex 9, France}

\author{C. Simon}
\affiliation{Laboratoire CRISMAT, UMR CNRS ENSICAEN, 1450 Caen, France}

\date{March 27, 2009}

\begin{abstract}
The Fe-based langasites are the first reported compounds presenting a magnetic ordering in this rich family, besides being well known for piezo-electric properties and optical activity. The structural, magnetic and dielectric properties of the Fe langasite compounds, with various substitution of non-magnetic cations, have been studied with x-ray and neutron diffraction, magnetostatic measurements, M$\mathrm{\ddot o}$ssbauer spectroscopy, and dielectric measurements. The title compounds (trigonal, space group $P321$) display a helical magnetic order with signatures of frustration below $T_{N}\approx$24-35 K where an anomaly of the dielectric permittivity is observed. The influence of the cationic substitutions and the nature of the magnetoelectric coupling is hereafter addressed.   
\end{abstract}


\maketitle

\section{Introduction\label{sec:intro}}

Multifunctional materials, in which coexist different physical properties (optical activity, laser properties, piezo- and/or ferroelectricity, ferroelasticity, magnetic ordering, etc…), are the subject of an intense research effort owing to their strong potential for various industrial applications. To date, the basic mechanisms at the origin of the coupling between these properties inside a single compound are still poorly understood. Experimental realizations of such materials are rare, and this is particularly the case in the field of multiferroics, i.e. compounds which display both ferroelectric and magnetic ordering \cite{Hill1}. Proper multiferroics are defined as ferroelectric compounds which undergo a magnetic ordering at lower temperature, as for example BiFeO$_{3}$\cite{BiFeO3}. In these compounds, there is a large temperature gap between the ferroelectric and magnetic transition, and the coupling between the two properties is generally weak. For improper multiferroics (for example TbMn$_{2}$O$_{5}$ \cite{TbMn2O5}), the electric polarization is induced by the onset of a complex magnetic order within a paraelectric and generally centrosymmetric phase. Such materials therefore present a naturally strong coupling between magnetization and electric polarizations thus giving the opportunity to study the microscopic mechanisms driving such couplings. 

The langasite family, the prototype of which is La$_3$Ga$_5$SiO$_{14}$, has been widely studied for the striking piezoelectric and non-linear optical properties of its members\cite{Mill2,Bohm3,Sato4,Iwataki5}, related to the non-centrosymmetric nature of its Ca$_3$Ga$_2$Ge$_4$O$_{14}$ structure type\cite{Mill6}. These compounds have generated a strong interest for applications in bulk acoustic waves (BAW) and surface acoustic waves (SAW) devices, as well as in non-linear optics and electro-optics\cite{Mill2,Stade7,Xin8}. 

\begin{figure}[b]
	\begin{center}
		\includegraphics[width=0.3\textwidth]{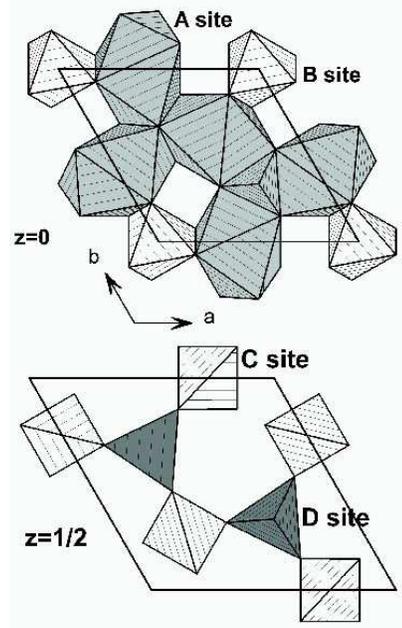}
		\caption{Polyhedral representation of the langasite structure with its four different cation sites on two layers.}
		\label{struct_1}
	\end{center}
\end{figure}

The langasite structure belongs to the trigonal non-centrosymmetric $P321$ space group. The general formula is A$_3$BC$_3$D$_2$O$_{14}$, thus containing four different cationic sites. The decahedral A site and the octahedral B site form a layer at z=0, while the two tetrahedral sites C and D are located on the plane z=1/2 (see Figure \ref{struct_1}) \cite{Mill2,Maksimov12}. 

Due to its atomic arrangement, this structure type is able to accommodate a large number of different cations with various sizes and valences, leading to a wide variety of isostructural compounds \cite{Mill2,Eysel9,Mill10}. Among these, some contain magnetic cation sublattices. Therefore, this family might provide interesting examples of coexisting magnetic order and optical or electrical properties. Recently, we have undertaken the general investigation of the magnetic langasites \cite{Bordet11}. In this article, we present the study of the magnetic and dielectric properties of several Fe-containing langasite compounds with formula A$_3$BFe$_3$D$_2$O$_{14}$, with A=Ba, Sr, Ca, B=Ta, Nb, Sb and D=Ge, Si.

\section{Experimental\label{sec:experimental}}

\subsection{Synthesis and structural characterization\label{synth}}

Powder samples were synthesized by solid state reactions. The initial components of the powder mixtures were high purity oxides or carbonates in stoichiometric quantity. The mixtures were ground, compacted and heated up to 1150 {-} 1250\degre C in Al$_2$O$_3$ crucibles in air for 7 hours. The pellets were heated at a rate ranging from 200\degre C/h to 300\degre C/h. The samples containing GeO$_2$ were kept at 1000\degre C for 2h before reaching the final temperature to prevent the loss of the relatively volatile germanium oxide. The synthesis conditions were chosen taking into account some properties of the initial and final compounds, such as the decomposition of the carbonates, the oxidation of Sb$_2$O$_3$, the fusion temperatures of GeO$_2$, SrO and of the final product. The products were analyzed with x-ray diffraction using a D8 Brucker diffractometer in transmission geometry with the CuK$\alpha$1 radiation, to check the presence of possible impurities. As some of the products showed minor impurities (max. 3 {\%}), they were ground, compacted again and annealed for 10-50h, up to three times in order to decrease the impurity content.

\begin{table}[b]
    \begin{center}
    \caption{\label{struc_BNFS} Refined structural parameters obtained from x-ray powder diffraction on BNFS with a D8 Brucker diffractometer. a. d. p. is the atomic displacement parameter. The agreement factor of the fit is given by  $R_{\rm Bragg}=3.2$~; $R_{\rm wp}=27.0$~; $\chi^2=1.5$ .}
    \vspace{2mm}
    \begin{tabular}{cccccc}
    \hline
    \hline
    Atom & Wyckoff & x & y & z & a.d.p.  \\
    \hline
    \hline
    Ba & 3e & 0.4339(2) & 0 & 0 & 1.04(1)  \\ 
    \hline
    Sb & 1a & 0 & 0 & 0 & 0.9(1)  \\
    \hline
    Fe & 3f & 0.7513(6) & 0 & 1/2 & 0.9(1)  \\
    \hline
    Si & 2d & 1/3 & 2/3 & 0.459(3) & 0.4(3)  \\
    \hline
    O1 & 2d & 2/3 & 1/3 & 0.774(6) & 1.4(3)  \\
    \hline
    O2 & 6g & 0.477(3) & 0.300(2) & 0.649(3) & 1.4(3)  \\
    \hline
    O3 & 6g & 0.209(2) & 0.098(2) & 0.236(3) & 1.4(3)  \\
    \hline
    \hline
    \multicolumn{3}{c}{a = 8.5227(2) \AA} & \multicolumn{3}{c}{c = 5.2353(1) \AA}  \\
    \hline
    \hline
    \end{tabular}
   \end{center}
\end{table}

\begin{table}[t]
    \begin{center}
    \caption{\label{refXall} Structural parameters and agreement factors of Fe-langasites as refined
from x-ray powder diffraction data at room temperature with a D8 Brucker diffractometer. }
    \begin{tabular}{cccccc}
    \hline
    \hline
    Langasite & a (\AA) & c (\AA) & $R_{\rm Bragg}$ & $R_{\rm wp}$ & $\chi^2$   \\
    \hline
    \hline
    Ba$_3$SbFe$_3$Ge$_2$O$_{14}$ & 8.6174(1) & 5.2708(1) & 3.9 & 22.7 & 3.2  \\
    \hline
    Ba$_3$TaFe$_3$Ge$_2$O$_{14}$ & 8.6166(1) & 5.2618(1) & 6.0 & 22.5 & 3.9  \\
    \hline
    Ba$_3$NbFe$_3$Ge$_2$O$_{14}$ & 8.6073(2) & 5.2686(2) & 6.1 & 22.9 & 2.5  \\
    \hline
    Ba$_3$SbFe$_3$Si$_2$O$_{14}$ & 8.5085(1) & 5.25024(8) & 3.1 & 20.2 & 2.8  \\
    \hline
    Ba$_3$TaFe$_3$Si$_2$O$_{14}$ & 8.5231(2) & 5.2354(2) & 13.5 & 33.6 & 2.2  \\
    \hline
    Ba$_3$NbFe$_3$Si$_2$O$_{14}$ & 8.5227(2) & 5.2353(1) & 3.2 & 27.0 & 1.5  \\
    \hline
    Sr$_3$SbFe$_3$Si$_2$O$_{14}$ & 8.2888(2) & 5.1445(2) & 5.8 & 30.6 & 2.1  \\
    \hline
    Sr$_3$TaFe$_3$Si$_2$O$_{14}$ & 8.2777(2) & 5.1283(1) & 4.3 & 27.0 & 2.0  \\
    \hline
    Sr$_3$NbFe$_3$Si$_2$O$_{14}$ & 8.2563(2) & 5.1306(1) & 3.2 & 27.6 & 1.6  \\
    \hline
    Ca$_3$SbFe$_3$Si$_2$O$_{14}$ & 8.1118(2) & 5.0570(2) & 5.7 & 26.0 & 2.0  \\
    \hline
    \end{tabular}
   \end{center}
\end{table}

Rietveld refinements with the program Fullprof\cite{Rodriguez15} were carried out on the final products diffractogramms. In all cases, the expected stoichiometry and the full occupation of all sites was found, without any sign of intersite substitution. As an example, the results for Ba$_3$NbFe$_3$Si$_2$O$_{14}$ (hereafter referred to as BNFS) are reported in Table \ref{struc_BNFS}. Table \ref{refXall} summarizes the principal structural parameters for the investigated samples.

Single crystals of Ba$_3$NbFe$_3$Si$_2$O$_{14}$ and Ba$_3$TaFe$_3$Si$_2$O$_{14}$ (BTFS) were grown by the floating-zone method in a Cyberstar image furnace, under a 99\% Ar + 1\% O$_2$ atmosphere, at a growth rate of 10~mm/h \cite{Lejay16}. This technique has the advantage of being fast and exempt of contamination from a crucible. The single crystals were cut out of the grown ingot with a size roughly of 3 mm in diameter and 8 mm in length for the neutron diffraction measurements. The crystallographic quality were checked by La$\mathrm{\ddot u}$e photographs. No impurity phase could be detected by x-ray powder diffraction of crushed crystal pieces. A small crystal piece of each compound was used for single crystal x-ray diffraction analysis using a Kappa ApexII Bruker diffractometer equipped with graphite monochromatized AgK$\alpha$ radiation. This confirmed the crystal quality and the langasite type structure and allowed to determine the handedness of each of these chiral crystals by use of anomalous scattering sensitivity.   

\begin{figure}[b]
	\begin{center}
		\includegraphics[width=0.4\textwidth]{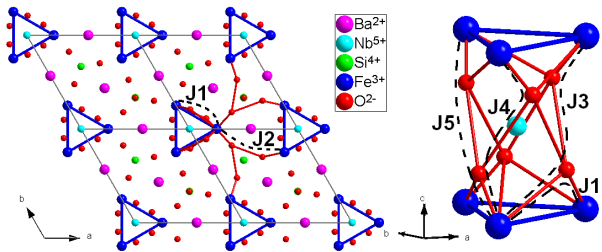}
		\caption{ Magnetic network of Fe$^{3+}$ cations projected along the $c$-axis (left), with the different intra-plane (left) or inter-plane (right) exchange paths (dashed lines).}
		\label{Feplan_3}
	\end{center}
\end{figure}

In all the compounds under study, the magnetic Fe$^{3+}$ cations form a peculiar array of triangles on a triangular lattice, in the plane at z=1/2 (figure \ref{Feplan_3}). Two consecutive planes are separated from each others by a layer of A$^{2+}$ and B$^{5+}$ cations. The Fe$^{3+}$ triangles are equilateral by symmetry. The distance between the Fe$^{3+}$ cations is 3.662(8) \AA~for BNFS. Two equivalent super-exchange interactions J1 between two nearest neighbor magnetic cations on one triangle are each mediated via an O3 anion. The Fe-O3-Fe angle 100.1(9)\degre and the Fe-O3 distances are 1.73(2) and 2.93(2) \AA~for BNFS. One Fe$^{3+}$ cation of a triangle interacts with four neighbors in the nearest triangles within the plane by the super-super-exchange paths J2 through two O2 anions forming an edge of a DO4 tetrahedron (figures \ref{struct_1} and \ref{Feplan_3}). The Fe-O2-O2 angles are 144(1)\degre and 140(1)\degre, and the Fe-O2 and O2-O2 distances are respectively 1.93(2) and 2.59(4) \AA~in BNFS. The shortest distance between two Fe$^{3+}$ cations from neighbouring planes is the c parameter (5.2353(1) \AA), with also two O2 oxygens in the exchange path forming an edge of the BO$_6$ octahedron. But one needs to distinguish three different inter-plane interactions, each one with multiple possible super-super-exchange paths. J4 is the interaction between a Fe$^{3+}$ cation and the cation just above or below in the adjacent plane, while J3 and J5 are the diagonal interactions between this Fe$^{3+}$ cation and the two other Fe$^{3+}$ cations of the neighbouring triangle (figure~\ref{Feplan_3}). It is worth noticing that due to the structural chirality, J3 and J5 are obtained through different exchange paths and expected not to be equal.

\subsection{Magnetization measurements\label{Magn}}

\begin{figure}[b]
	\begin{center}
		\includegraphics[width=0.4\textwidth]{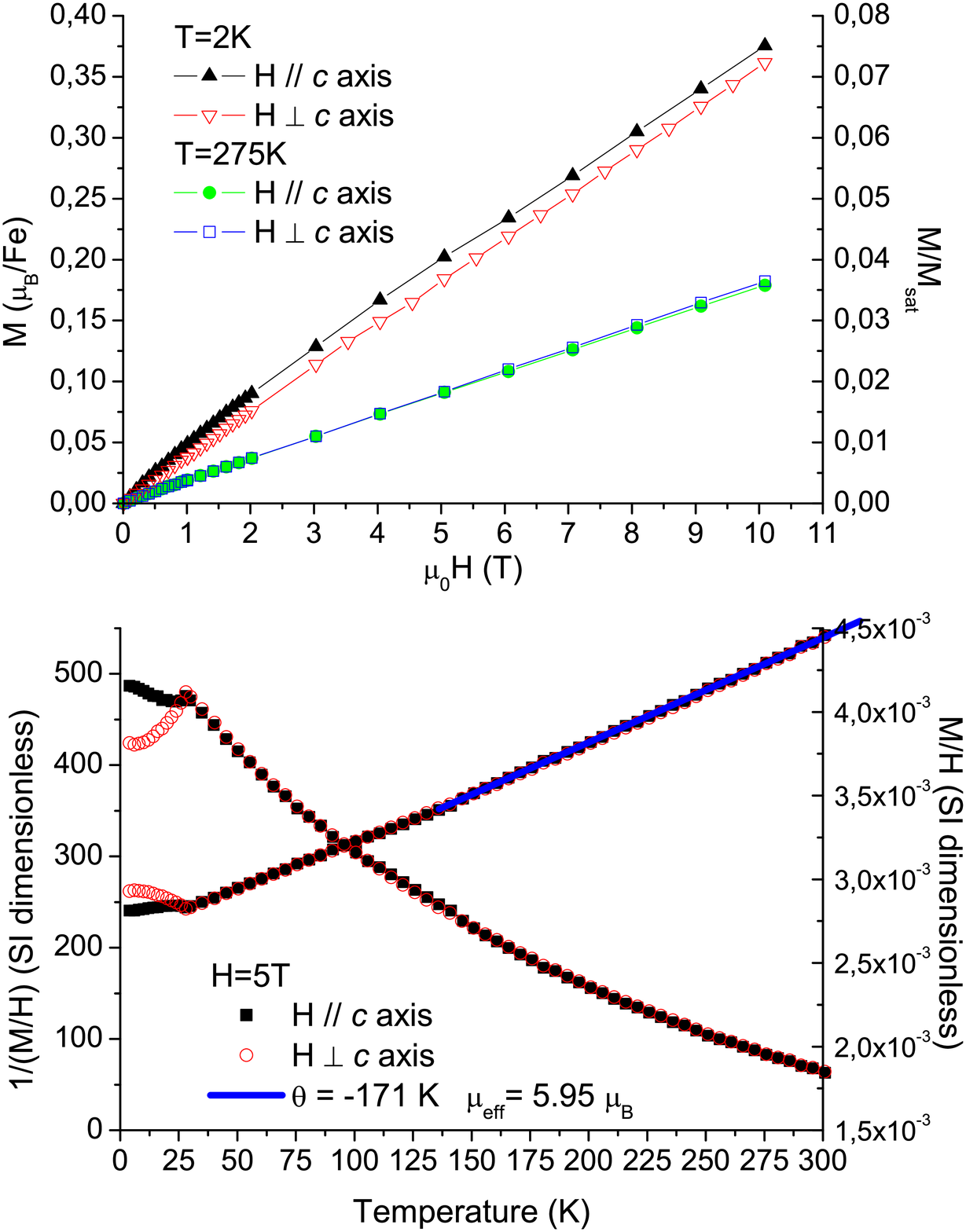}
		\caption{ Magnetization measurements on a Ba$_3$NbFe$_3$Si$_2$O$_{14}$ single crystal under a magnetic field applied parallel and perpendicular to the $c$-axis. Top: Magnetic isotherms at 2 K and 275 K. Bottom: Thermal variation of the $M/H$ and its inverse under an applied magnetic field of 5 T. The continuous line represents the linear fit of $H/M$ in a Curie-Weiss model.}
		\label{suscep_4}
	\end{center}
\end{figure}

The magnetostatic properties of BNFS and BTFS single crystals and of all the powder title compounds were investigated under magnetic fields up to 5 T in the temperature range from 2 K to 300 K with a commercial Quantum Design MPMS SQUID magnetometer, and up to 10 T from 1.6 K to 300 K with a purpose-built magnetometer using the anti-Helmholtz two-coil axial extraction method. Figure \ref{suscep_4} illustrates for BNFS the thermal evolution of the magnetization divided by the magnetic field (i.e. equal to the linear susceptibility $\chi$ for temperatures greater than $\approx$ 26 K), and of its inverse, measured under a magnetic field of 5 T applied both perpendicular to the $c$-axis and parallel to it. A cusp in the $M/H$ measurements suggested the onset of a magnetic transition at 26 K, further corroborated by a sharp peak in the specific heat \cite{BNFS_JMMM}. Above this temperature, no significant anisotropy is observed within the accuracy of the measurements, which is expected for Fe$^{3+}$ ions without orbital contribution. Below the magnetic transition, whereas $M/H$ is almost constant for H parallel to the $c$ axis, the in-plane susceptibility drops, although not to zero as would be the case for a canonical antiferromagnetic transition. This implies a more complex magnetic arrangement. This is also illustrated on the magnetization curves $M(H)$ shown on figure \ref{suscep_4}. At high temperature (275 K), the magnetization is the same for the two orientations of the applied field. On the contrary, at low temperature (2 K), an anisotropy of the magnetostatic properties comes out: the magnetization measured with an applied field parallel to the $c$ axis is no longer linear and shows a slight curvature at small fields. This curvature would indicate the rise of a small component along the $c$ axis. At high temperatures (150 K to 300 K), a linear fit of the inverse of $M/H$ yields, in a Curie-Weiss model $\chi=C/(T-\theta)$, an effective moment $\mu_{\rm eff}=g\sqrt{S(S+1)}$=5.95 $\mu_B$, compatible with the value of the Fe$^{3+}$ free ion ($3d^{5}$, $J$=5/2, $\mu_{\rm eff}$=5.92 $\mu_B$, $M_{sat}$=5 $\mu_B$), and a negative Curie-Weiss temperature of -171 K, characteristic of antiferromagnetic interactions. 

\begin{figure}[t]
	\begin{center}
		\includegraphics[width=0.35\textwidth]{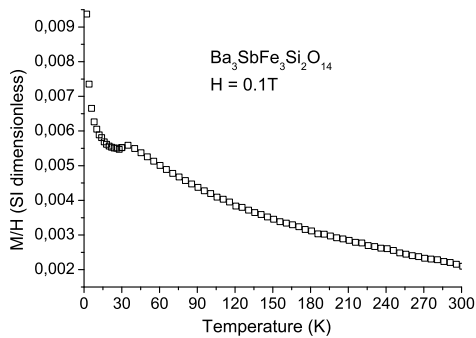}
		\caption{ Thermal variation of the $M/H$ of a powder sample of Ba$_3$SbFe$_3$Si$_2$O$_{14}$ under an applied magnetic field of 0.1 T.}
		\label{susc_BSFS}
	\end{center}
\end{figure}

\begin{table}[t]
    \begin{center}
    \caption{\label{magnet} Magnetic parameters obtained from magnetization measurements of Fe-langasites: N\'eel temperatures and Curie-Weiss temperatures.}
    \vspace{2mm}
    \begin{tabular}{ccc}
    \hline
    \hline
    Langasite & T$_N$ (K) & $\theta$ (K) \\
    \hline
    \hline
    Ba$_3$NbFe$_3$Si$_2$O$_{14}$ & 26 & -173  \\
    \hline
    Ba$_3$NbFe$_3$Ge$_2$O$_{14}$ & 24 & -177  \\
    \hline
    Ba$_3$TaFe$_3$Si$_2$O$_{14}$ & 28 & -160  \\ 
    \hline
    Ba$_3$TaFe$_3$Ge$_2$O$_{14}$ & 24 & -165  \\ 
    \hline
    Ba$_3$SbFe$_3$Si$_2$O$_{14}$ & 35 & -114  \\ 
    \hline
    Ba$_3$SbFe$_3$Ge$_2$O$_{14}$ & 34 & -111  \\ 
    \hline
    Sr$_3$NbFe$_3$Si$_2$O$_{14}$ & 26 & -242   \\ 
    \hline
    Sr$_3$SbFe$_3$Si$_2$O$_{14}$ & 36 & -128   \\ 
    \hline
    \hline
    \end{tabular}
   \end{center}
\end{table} 

The magnetic behavior of a BTFS crystal was found essentially similar to that of BNFS, and this was also the case for all the investigated compounds prepared as powders : all of them undergo an antiferromagnetic ordering transition in the 24-35 K range. As an example, the magnetic susceptibility for the Ba$_3$SbFe$_3$Si$_2$O$_{14}$ powder compound is shown in Figure \ref{susc_BSFS} where the magnetic transition is clearly visible at $T_{N}$=35 K. The values of $T_{N}$, $\theta$ and Fe$^{3+}$ magnetic effective moments obtained in a Curie-Weiss fit of the inverse magnetic susceptibilities are summarized in Table \ref{magnet}. All compounds, whatever the cations on sites A, B and D,  present a N\'eel temperature around 25 K, except those containing  Sb$^{5+}$ cations on site B. These latter display a magnetic transition around 35 K, and at the same time a noticeably smaller absolute value of the Curie-Weiss temperature. In order to understand the microscopic origin of this difference, and more generally to further our understanding of the magnetic order in these compounds, microscopic probes such as neutron scattering and M\"ossbauer spectroscopy were used.

\subsection{Neutron diffraction study\label{Neutron}}

\begin{figure*}[t]
	\begin{center}
		\includegraphics[width=0.9\textwidth]{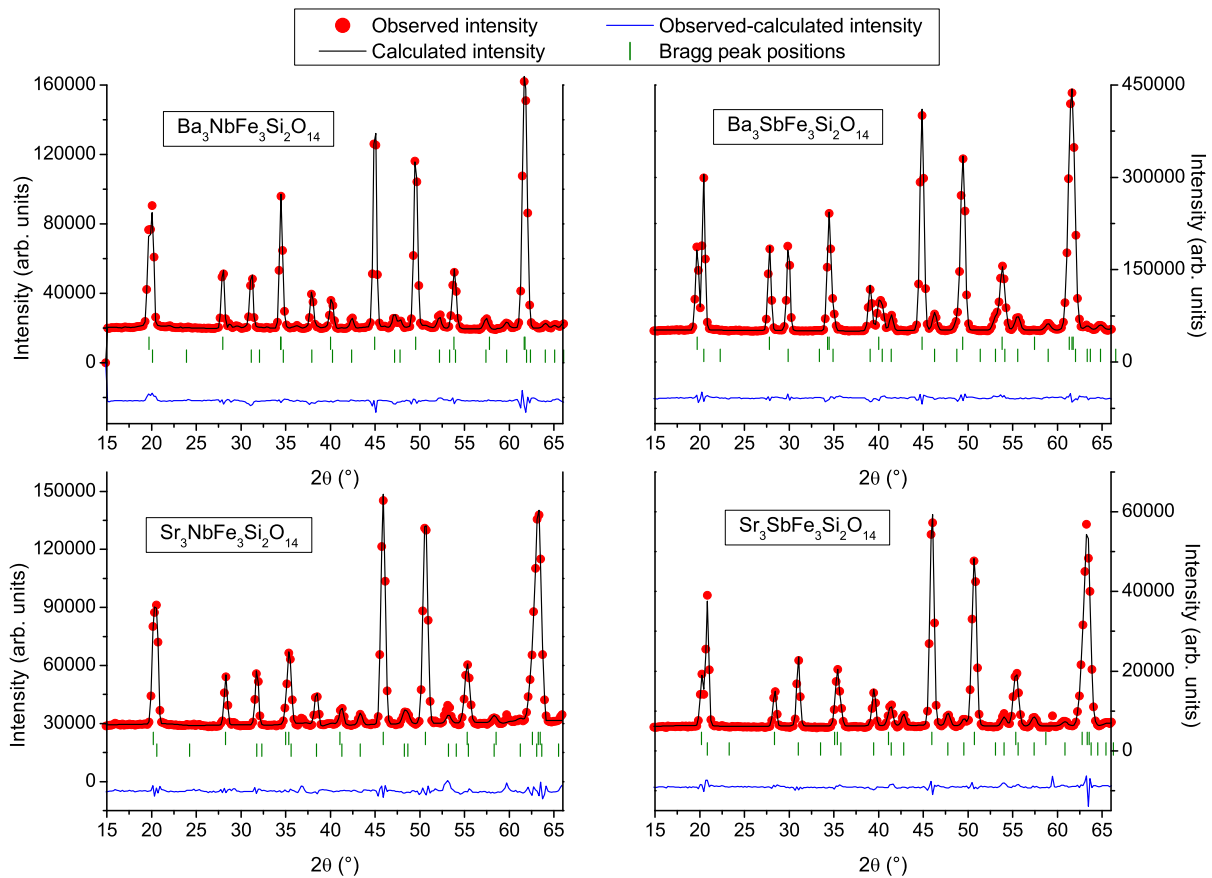}
		\caption{Rietveld plot for Ba$_3$NbFe$_3$Si$_2$O$_{14}$ (top left), Ba$_3$SbFe$_3$Si$_2$O$_{14}$ (top right), Sr$_3$NbFe$_3$Si$_2$O$_{14}$ (bottom left) and Sr$_3$SbFe$_3$Si$_2$O$_{14}$ (bottom right) neutron powder data at 1.8 K (1.5 K for BNFS) from D1B, including nuclear  and magnetic phases.}
		\label{D1B}
	\end{center}
\end{figure*}

\begin{figure}[t]
	\begin{center}
		\includegraphics[width=0.4\textwidth]{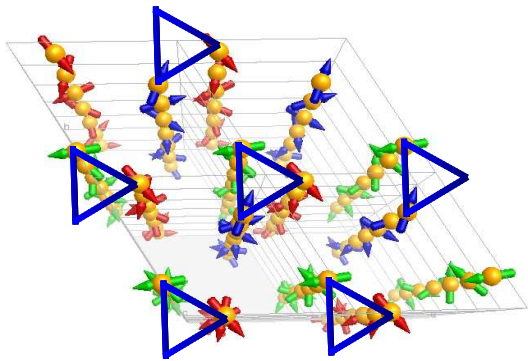}
		\caption{Magnetic structure of the Fe-langasites: within each triangle of the ($a$,$b$) plane, the Fe$^{3+}$ magnetic moments are rotated by 2$\pi$/3 from each other. The global phase factor, which cannot be determined by neutron diffraction, was arbitrary chosen equal to zero in this representation (one moment pointing towards the center of the triangles). Along the $c$-axis, the helix propagates with a period of $1/\tau$ in cell unit.}
		\label{Struc_Mag}
	\end{center}
\end{figure}

Neutron powder diffraction (NPD) measurements were performed at the Institut La\"ue Langevin on samples of the compounds Ba$_3$NbFe$_3$Si$_2$O$_{14}$ (BNFS), Ba$_3$SbFe$_3$Si$_2$O$_{14}$ (BSFS), Sr$_3$NbFe$_3$Si$_2$O$_{14}$ (SNFS) and Sr$_3$SbFe$_3$Si$_2$O$_{14}$ (SSFS) using the D1B instrument ($\lambda$=2.52 \AA). These four compounds were chosen in order to track, via the structural and magnetic arrangements determination, the influence of large cell parameter variation (Sr versus Ba) and the microscopic origin of the different transition temperatures (Nb versus  Sb). A BNFS single crystal was also studied on the D15 diffractometer of the Institut La\"ue Langevin ($\lambda$=1.1743 \AA). The thermal evolution of the diffraction patterns were measured on D1B using an orange cryostat by increasing the temperature at a rate of 0.42 K/min with 75 scans of 10 minutes each. In order to obtain sufficient statistics at 30 K (above the magnetic transition), 20 K (below the magnetic transition) 1.8 K (for BSFS, SNFS and SSFS) and 1.5 K (the lowest reachable temperature, for BNFS), 6 scans of 20 minutes were summed at each temperature. The magnetic structure was solved from refinement of the D15 and D1B diffractograms measured on BNFS \cite{BNFS_JMMM, BNFS_PRL}. The magnetic structures of the three other compounds were refined using the same ordering scheme (see figure~\ref{D1B}). The magnetic peaks could be indexed using an incommensurate propagation vector (0, 0, $\tau$), with $\tau$ ranging from 0.14 for SNFS to 0.19 for BSFS, and the magnetic structure was found to be a helical spin arrangement propagating along the $c$ axis from equal moments lying in the ($a$,$b$) plane at 120\degre from each other within each triangle (see figure~\ref{Struc_Mag}). This 120\degre arrangement results from the usual compromise of frustrated Heisenberg spins on a triangle-based lattice. All spins rotate around $c$ by $\pm2\pi\tau$ between to consecutive Fe$^{3+}$ layers. It is worth noting that a helical magnetic order often results either from the presence of frustration of interactions in the helix direction\cite{Yoshimori}, or from Dzyaloshinskii-Moriya interactions in a ferromagnetic structure\cite{Dzyaloshinskii}. In the present case, it was shown that the helical stucture is generated by the twist in the magnetic super-super-exchange paths from plane to plane. Furthermore, the in-plane 120\degre magnetic arrangement and the helical propagation both possess a definite sense of rotation defined as chirality, which is related to the structural chirality. A thorough investigation of this interesting aspect has been previously reported in ref. \onlinecite{BNFS_PRL} and will not be discussed further here.

\begin{table}[t]
    \begin{center}
    \caption{\label{Table_D1B} Refined lattice parameters, magnetic moment value, overall isotropic atomic displacement parameter, $z$-component of the propagation vector and goodness of the fit parameters obtained from neutron powder diffraction data of Fe-langasites at 1.8 K (1.5 K for BNFS) on D1B.}
    \vspace{2mm}
    \begin{tabular}{ccccccc}
    \hline
    \hline
    Langasite & a (\AA) & $m$ ($\mu_B$) & $B_{mag}$ &  $\tau$ & R$_{\rm Bragg}$ &  $\chi^2$  \\
     & \textit{c (\AA)} & & & & $R_{mag}$ &  \\
    \hline
    \hline
    BNFS & 8.5026(6) & 4.04(5) & 3.9(9) & 0.1429(2) & 0.7  & 28.0    \\
     & \textit{5.2136(7)} & & & & \textit{5.8} \\
    \hline
    SNFS & 8.297(1) & 3.87(9) & 5(2) & 0.1398(3) & 0.9 & 35.5   \\
     & \textit{5.155(1)}  & & & & \textit{9.2}   \\
    \hline
    BSFS & 8.5048(5) & 4.26(3) & 2.1(4) & 0.1957(1) & 1.2 & 44.3   \\
     & \textit{5.2449(4)}  & & &  & \textit{4.8}   \\
    \hline
    SSFS & 8.2896(9) & 4.31(5) & 4.6(7) & 0.1769(2) & 1.8 & 20.3  \\
     & \textit{5.140(1)}  & & &  & \textit{14.2}   \\
    \hline
    \hline
    \end{tabular}
   \end{center}
\end{table}

\begin{figure}[t]
	\begin{center}
		\includegraphics[width=0.4\textwidth]{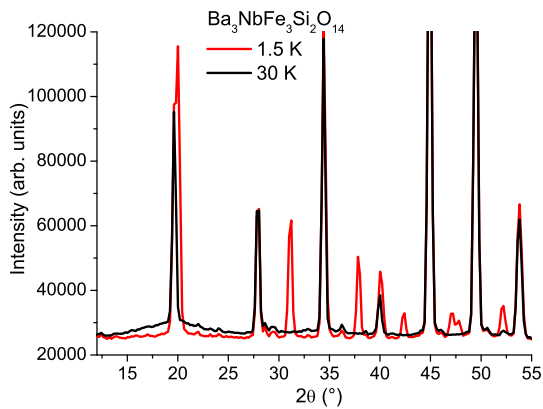}
		\caption{ Powder neutron diffraction patterns of the compound Ba$_3$NbFe$_3$Si$_2$O$_{14}$, below the magnetic transition, at 1.5 K, and above the magnetic transition, at 30 K.}
		\label{BNFS_diff}
	\end{center}
\end{figure}

The main results of the NPD refinements carried out for the four samples at 1.8 K are summarized in Table \ref{Table_D1B}. The magnetic moment values are close to 4 $\mu_B$ and they are smaller for Nb compounds than for Sb ones. This is markedly smaller than the expected value for a saturated Fe$^{3+}$ magnetic moment (5 $\mu_B$). An explanation for this behavior could be the strongly covalent nature of the Fe-O bonds for Fe$^{3+}$ cations in tetrahedral coordination, which is known\cite{Fe3O4} to favor spin transfer towards the oxygen orbitals and therefore decrease the apparent magnetic moment on the Fe$^{3+}$ cations. Indeed, in order to obtain accurate refinements of the NPD data, we had to introduce an ad-hoc "magnetic Debye-Waller factor", similar to an overall isotropic atomic displacement term.  The refined values of the corresponding $B_{mag}$ parameters are given in Table \ref{Table_D1B}. The effect of this term can be viewed qualitatively as a modification of the magnetic form factor with increased spin density at larger distances from the center of the electron cloud, as would be expected in the presence of spin transfer to the oxygen orbitals. Another qualitative information is given by the diffractograms measured at temperatures up to 20 K above the magnetic transition, where some diffuse scattering centered on the first nuclear peak is visible. This signal is characteristic of the onset of short range magnetic correlations  in the ($a$, $b$) planes. The width of the signal decreases, i.e. the correlation length of this short-range order increases, down to the magnetic transition where it disappears at the expense of the magnetic Bragg peaks characteristic of the 3-dimensional long-range order (see figure~\ref{BNFS_diff}).

As seen in Table \ref{Table_D1B}, the $\tau$ values of the antimony compounds are higher (between 0.177 and 0.196) than those of the niobium compounds (around 0.14). The $\tau$ values appear also slightly larger for the barium compounds than for the strontium ones, especially in the Sb compounds. However, the $\tau$ values are much less sensitive to the Sr/Ba substitution than to the Sb/Nb one, although the former have a much larger effect on the lattice parameter due to the big difference of their ionic radius (1.56 \AA\ for Ba versus 1.4 \AA\ for Sr).  After examination of the refined structures of the four compounds, and especially focusing on the lengths and bonds angle of the magnetic exchange paths, no systematic trends could be extracted from the data that would explain the peculiarity of the Sb based compounds magnetic properties on pure structural grounds. The influence of the Sb cation in the Fe-langasites is therefore more probably ascribable to the ion electronic state than to induced structural distortion. We will come back to this aspect in the discussion.

\subsection{M$\mathrm{\ddot o}$ssbauer measurements\label{Moss}}

\begin{figure}[t]
	\begin{center}
		\includegraphics[width=0.4\textwidth]{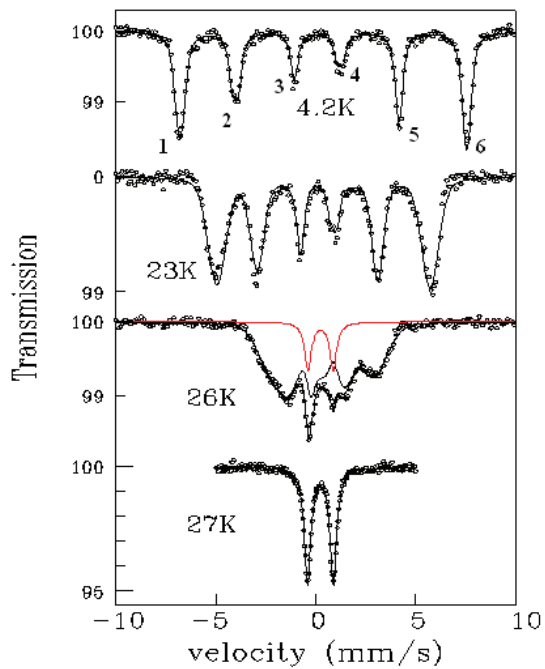}
		\caption{ $^{57}$Fe M$\mathrm{\ddot o}$ssbauer absorption spectra in Ba$_3$NbFe$_3$Si$_2$O$_{14}$ at selected temperatures. The spectra at 4.2, 23 and 26 K (circles) are fitted (black lines) to the helical magnetic structure model determined by neutron scattering. At 26 K, a quadrupolar subspectrum contribution (represented in red) overlapping the magnetic signal suggests that the magnetic transition is first order.}
		\label{Moss_5}
	\end{center}
\end{figure}

M$\mathrm{\ddot o}$ssbauer absorption spectra on the isotope $^{57}$Fe ($E_0$ = 14.4 keV) were recorded for the BNFS compound at room temperature and in the temperature range 4.2 to 30 K, in zero external magnetic field 
(see figure \ref{Moss_5}). At 27 K and above, the spectrum consists of a quadrupolar hyperfine doublet characteristic of the paramagnetic phase. At 27 K, the isomer shift with respect to the $\alpha$-Fe reference is 
$\delta$ = 0.35(1) mm/s (1 mm/s corresponds to 11 MHz) and the quadrupolar hyperfine splitting is $\Delta E_{Q} = \left|eQV_{ZZ}\right|/2$ = 1.29(1) mm/s, where $Q$ is the quadrupole moment of the nuclear excited state and 
$V_{ZZ}$ is the principal component of the electric field gradient (EFG) tensor at the $^{57}$Fe nucleus site. These values are typical of a Fe$^{3+}$ ion, although the $\Delta E_{Q}$ value is somewhat larger than generally 
found in ferric insulators, indicating a strongly distorted Fe site, compared to a perfect tetrahedral environment, in agreement with crystal analysis. The point symmetry at the Fe site is low (a two-fold axis along the crystal $a$ 
axis), and thus the directions of the EFG tensor axes cannot be determined from symmetry considerations alone.

Below 27 K, a magnetic hyperfine structure appears, with 6-line characteristic spectra showing the presence of a magnetic hyperfine field acting on the nuclear moment. This is the fingerprint of a long range magnetic order. The 
4.2 K spectrum (see figure \ref{Moss_5}) corresponds to a single saturated hyperfine field $H_{hf}$ = 440(2) kOe; considering the hyperfine proportionality constant typical for Fe$^{3+}$ insulators of 110 kOe/$\mu_{\rm B}$, this yields a unique Fe$^{3+}$ magnetic moment value of 4 $\mu_{\rm B}$, in agreement with the neutron derived value. 
Close inspection of the 4.2 K spectrum reveals slight inhomogeneous line broadenings, most clearly seen when comparing the widths of the two central lines (labelled 3 and 4 in figure \ref{Moss_5}) and of the two intermediate lines (2 and 5). This spectral effect is due to a distribution of one or more hyperfine parameters. 
The magnetic structure determined by neutron diffraction being an incommensurate transverse spiral with propagation vector along $c$, then the angle between the hyperfine field, proportional to the Fe$^{3+}$ moment, and the principal axis OZ of the EFG tensor, linked to the crystal axes, shows a distribution. The second order shifts of the hyperfine energies due to the quadrupolar coupling then lead to a broadening which is different for each 
line \cite{lebeugle}. The line fits in the magnetic phase in Fig.\ref{Moss_5} were performed using this model, the key parameter determining the size of the broadenings being the angle $\Theta$ between OZ and the propagation vector, which is found to be $\Theta \simeq 36^\circ$. 
On heating, the lines become broader and the overall magnetic splitting decreases, indicating that the mean hyperfine field decreases and that a distribution in hyperfine field values appears. The broadenings are significant near $T_{N}$ (see the spectra at 23 and 26 K in figure \ref{Moss_5}). At 26 K, a small intensity paramagnetic doublet (13\% relative weight) is present together with the magnetic order spectrum. This coexistence of paramagnetic and magnetically ordered regions near $T_N$ is usually associated with a certain degree of first order for the magnetic transition. At 27 K, the sample as a whole is paramagnetic.

\subsection{High resolution X-ray diffraction\label{ESRF}}

\begin{figure}[b]
	\begin{center}
		\includegraphics[width=0.45\textwidth]{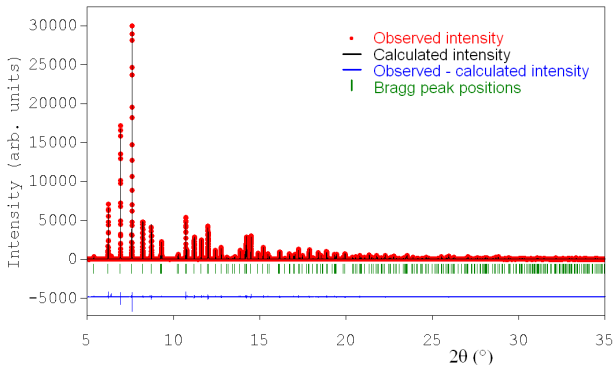}
		\caption{ Rietveld plot for Ba$_3$SbFe$_3$Si$_2$O$_{14}$ x-ray powder data at 250 K from ID31; R$_{Bragg}$ = 2.58, R$_{\rm wp}=23.4$, $\chi^2$ = 3.6.}
		\label{250K_ESRF}
	\end{center}
\end{figure}

\begin{table}[t]
    \begin{center}
    \caption{\label{BSFS_ID31} Refined structural parameters obtained from high resolution x-ray powder diffraction on BSFS on ID31, at 250 K (R$_{\rm Bragg}=2.58$~; R$_{\rm wp}=23.4$~; $\chi^2=3.6$) and \textit{10~K} ($R_{Bragg}=2.35$~; $R_{wp}=22.0$~; $\chi^2=0.95$). a.d.p. and U are respectively the atomic displacement parameter and the quadratic term of the Gaussian peak width.}
    \vspace{2mm}
    \begin{tabular}{cccccc}
    \hline
    \hline
    \multicolumn{6}{c}{Structural parameters refined at} \\ 
    \multicolumn{6}{c}{250~K}  \\
    \multicolumn{6}{c}{\textit{10~K}}  \\
    \hline
    \hline
    Atom & Wyckoff & x & y & z & a.d.p.  \\
    \hline
    \hline
    Ba & 3e & 0.43747(7) & 0 & 0 & 1.29(1)  \\ 
     &  & \textit{0.43686(6)} &  &  & \textit{1.000(8)}  \\ 
    \hline
    Sb & 1a & 0 & 0 & 0 & 1.07(2)  \\
    & & & & & \textit{0.92(2)}  \\ 
    \hline
    Fe & 3f & 0.7527(2) & 0 & 1/2 & 1.27(3)  \\
     &  & \textit{0.7523(1)} &  &  & \textit{1.11(2)}  \\
    \hline
    Si & 2d & 1/3 & 2/3 & 0.481(1) & 1.22(7)  \\
     &  &  &  & \textit{0.479(1)} & \textit{1.00(7)}  \\
    \hline
    O1 & 2d & 2/3 & 1/3 & 0.790(2) & 1.39(8)  \\
     &  &  &  & \textit{0.783(2)} & \textit{1.02(6)}  \\
    \hline
    O2 & 6g & 0.4739(9) & 0.2966(8) & 0.649(1) & 1.39(8)  \\
     &  & \textit{0.4742(9)} & \textit{0.2956(7)} & \textit{0.647(1)} & \textit{1.02(6)}  \\
    \hline
    O3 & 6g & 0.2174(7) & 0.1001(6) & 0.219(1) & 1.39(8)  \\
     &  & \textit{0.2161(7)} & \textit{0.0986(6)} & \textit{0.223(1)} & \textit{1.02(6)}  \\
    \hline
    \hline
    \multicolumn{2}{c}{a (\AA)} & \multicolumn{2}{c}{c (\AA)} & \multicolumn{2}{c}{U} \\
    \hline
    \multicolumn{2}{c}{8.51009(1)} & \multicolumn{2}{c}{5.251660(7)} & \multicolumn{2}{c}{0.0071(3)} \\
    \multicolumn{2}{c}{\textit{8.50665(1)}} & \multicolumn{2}{c}{\textit{5.247587(8)}} & \multicolumn{2}{c}{\textit{0.0122(4)}} \\
    \hline
    \hline
    \end{tabular}
   \end{center}
\end{table}

\begin{figure*}[t]
	 \begin{center}
		 \includegraphics[width=0.85\textwidth]{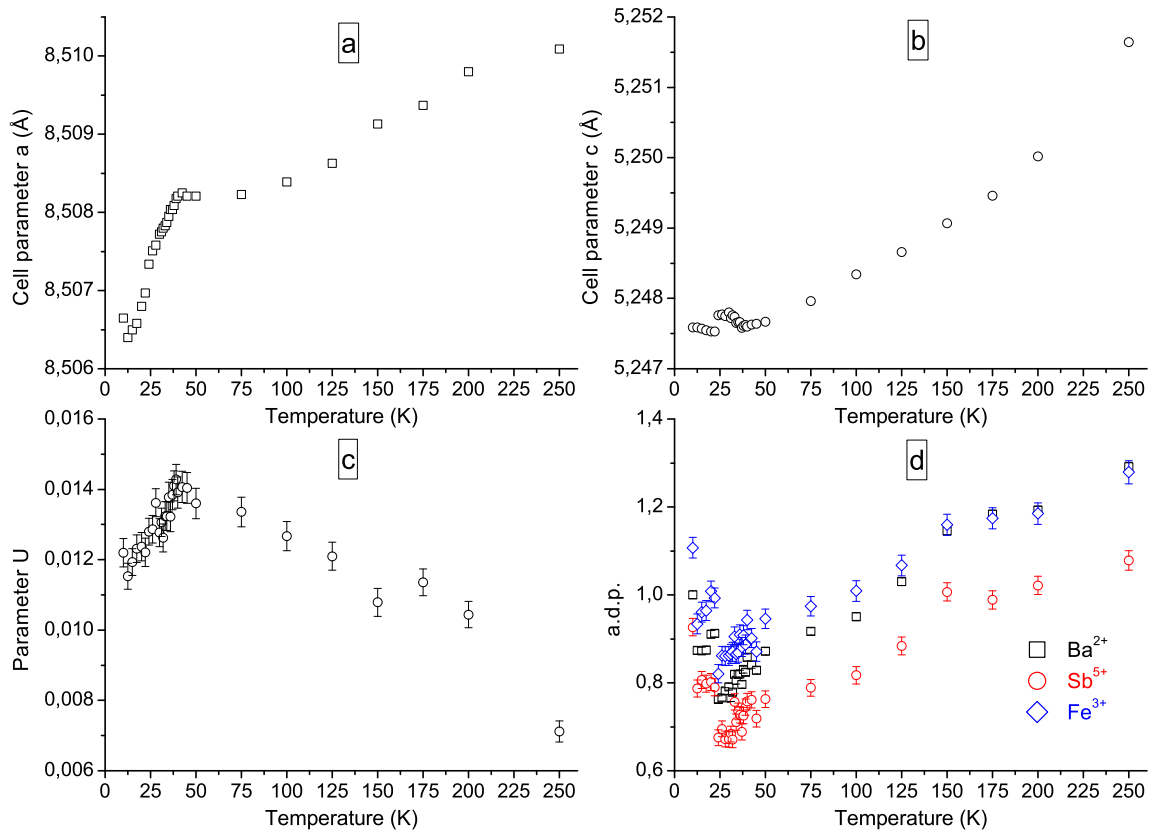}
		 		 \caption{ Thermal variation of some refined parameters from ID31 x-ray powder diffraction data on Ba$_3$SbFe$_3$Si$_2$O$_{14}$: a) cell parameter a ($\AA$). b) cell parameter c ($\AA$). c) quadratic term U of the Gaussian peak width function. d) a.d.p.'s of Ba$^{2+}$, Sb$^{5+}$ and Fe$^{3+}$ cations.}
		 \label{adp_ESRF}
	 \end{center}
\end{figure*}

 The remarkable magnetic properties of the Fe-langasites underlined in the previous sections may also have consequences on their dielectric behaviour. Indeed, the non-centrosymetric $P321$ space group symmetry of the langasites does not allow the existence of an electric polarization. However, we have established that the Fe-langasite compounds undergo a transition towards a magnetic helical order whose symmetries are compatible with ferroelectricity: preservation of the 3-fold axis and loss of the 2-fold axes. It is therefore interesting to determine whether a structural phase transition and a lowering of symmetry authorizing a ferroelectric order appear in the temperature region of the magnetic transition. The D1B data have not revealed any such modification in the four samples studied. However the limited Q-range and spatial resolution make this instrument moderately suitable for the detection of subtle structural anomalies. In order to follow the structure as function of temperature with a much higher sensitivity, we have used high resolution synchrotron powder diffraction at the ID31 beam line of the ESRF, equipped with a Ge(111) multi-analyzer stage. This experiment was performed on the Ba$_3$SbFe$_3$Si$_2$O$_{14}$ compound. The sample contained in a 1 mm diameter quartz capillary was placed inside a helium flow horizontal cryostat allowing sample rotation. Data collections were carried out in the range 2$\theta$=5\degre to 35\degre at 26 temperatures between 250 K and 10 K at an incident energy of 30 keV ($\lambda$=0.39816(2) \AA). The diffractograms were analyzed by Rietveld refinement using the Fullprof program. The background was interpolated and the Bragg reflection shapes were described using a Thomson-Cox-Hastings profile function and a uniaxial description of anisotropic peak broadening about the [001] direction. All positional parameters in the $P321$ space group description were refined, and all atoms were given isotropic atomic displacement parameters (a.d.p.), those of all oxygen atoms being constrained to be equal. The Rietveld plot for the 250 K data refinement is shown in figure~\ref{250K_ESRF}.

The refined parameters at 250 K and 10 K are given in table~\ref{BSFS_ID31} and do not exhibit any anomalous difference, the only marked variations being the expected cell contraction and the decrease of the a.d.p.'s with temperature. Figure~\ref{adp_ESRF} a, b, c and d display the temperature evolution of the cell parameters, quadratic term U of the Gaussian peak width function, and a.d.p.'s of the cations. An obvious anomaly is detected in the a-parameter variation below 35 K (i.e. the magnetic ordering transition temperature for this compound). However the amplitude of the anomaly is quite small since the cell parameter change between 40 K and 10 K is only about 0.0015 \AA. The U parameter depends on the instrumental resolution (which does not vary with temperature) and the presence of strain in the sample. This parameter, related to Gaussian strain, increases on cooling and then starts decreasing below about 40 K. This could be related to the building up of strain in the compound followed by a relaxation at lower temperature due to a structural reorganisation associated to a phase transition. In this context, the increase of the cation isotropic a.d.p.'s could be due to the accommodation of atomic displacements associated to this structural transition.

If a structural phase transition coupled to the onset of magnetic order was present, then the symmetry of the low temperature phase should keep the three-fold axis and loose the two-fold axis. The expected symmetry would then be $P3$, which allows an electric polarization. In this space group, the silicon and each of the oxygen sites are split into two positions. Refinements of the 10 K synchrotron data with the $P3$ symmetry were attempted, but did not lead to conclusive results. However, the departure from the $P321$ symmetry may be very small leading to hardly detectable atomic displacements. As a conclusion, the results of this temperature dependent high resolution structural study might indicate the presence of a structural phase transition coupled to the onset of magnetic order. Single crystal x-ray diffraction experiments are underway to determine the nature of the low temperature phase.

\subsection{Dielectric measurements\label{Dielec}}

In order to test the possible influence of the magnetic ordering on the dielectric properties of Fe-langasites, dielectric measurements were performed at the N\'eel Institute by measuring the complex impedance of our samples using a commercial HP 4284A LCR-meter. The purpose-built experimental setup consists of a sample holder protected by a dewar for a direct immersion in liquid helium of an helium container. Four coaxial cables linked to the electronic devices allow the complex impedance measurement. A model of capacitor and resistance in parallel was selected to extract the dielectric permittivity constant proportionnal to the capacitance of the sample, designed to approach the ideal geometry of a parallel plate capacitor. The metallic electrodes are realized by sputtering with a silver target. Systematic check of the amplitude and frequency dependence were undertaken to establish the better conditions of measurements ruling out extrinsic effects. The chosen conditions were finally an amplitude voltage of 1V and a frequency within the range of 10-100 kHz. The measurements under magnetic field were performed at the CRISMAT laboratory, with a similar experimental setup designed to work in a commercial Quantum Design PPMS and allowing to reach magnetic fields of 9 T.

The powder samples of BNFS, BTFS and BSFS were prepared by pressing thin pellets which were annealed at 1100\degre C during 78 hours. Single crystal samples of BNFS and BTFS were oriented and cut within the rods obtained in the image furnace in order to have two different orientations for the applied electric field: along the $c$ axis and along the $a$* axis.

\begin{figure}[b]
	 \begin{center}
		 \includegraphics[width=0.5\textwidth]{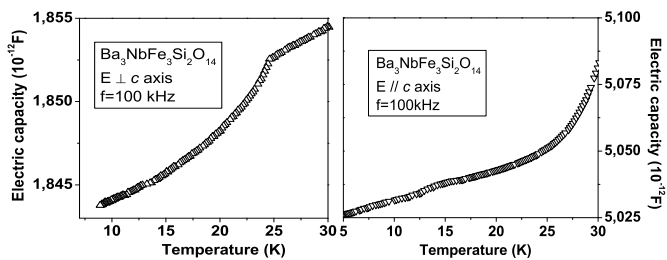}
		 \caption{ Thermal variation of the electric capacity (proportional to the dielectric permittivity) of BNFS single crystal for an electric field applied perpendicular to the $c$ axis (left) and parallel to it (right).}
		 \label{C(T)}
	 \end{center}
\end{figure}

\begin{figure}[t]
	 \begin{center}
		 \includegraphics[width=0.3\textwidth]{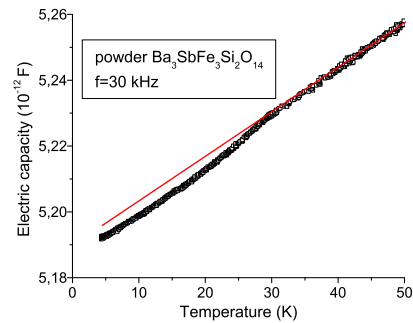}
		 \caption{Thermal variation of the electric capacity (proportional to the dielectric permittivity) of a pellet of BSFS powder. The red line is a linear extrapolation of the data above 30 K, shown to emphasize the cusp.}
		 \label{C(T)BSFS}
	 \end{center}
\end{figure}

The thermal variation of the dielectric permittivity constant of all the powder samples shows a small cusp near the temperature associated with the magnetic transition (24 K for BNFS and BTFS, 30 K for BSFS). This cusp is a contribution from the dielectric permittivity within the ($a$,$b$) plane. It is indeed clearly visible in the single crystal samples with the electric field applied along the $a$* axis (also reported in ref.\cite{zhou}), while it is absent for the other orientation perpendicular to the ($a$,$b$) plane which only shows a broad variation at higher temperature without any slope discontinuity (see figure~\ref{C(T)} for BNFS). For a powder sample (see BSFS measurement on figure~\ref{C(T)BSFS}), the observed cusp is obviously smaller than for a single crystal, as crystallites of every orientation, even those not presenting this effect, contribute to the signal.

\begin{figure}[t]
	 \begin{center}
		 \includegraphics[width=0.45\textwidth]{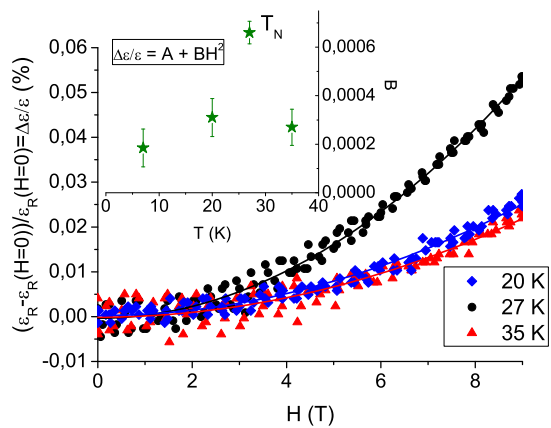}
		 \caption{ Magnetic dependance of the relative variation of the BNFS single crystal dielectric permittivity, for several temperatures. Both magnetic and electric fields are applied perpendicular to the $c$ axis. The lines are parabolic fits. The inset shows the quadratic coefficient versus temperature.}
		 \label{Eps(H)}
	 \end{center}
\end{figure}

The variation of the relative increase of the dielectric permittivity as a function of the magnetic field applied in the direction of the electric field, perpendicular to the $c$ axis, is illustrated for different temperatures by the figure~\ref{Eps(H)}. It shows mainly a quadratic dependence with the field. As the dielectric permittivity is the second partial derivative of the free energy with respect to the electric field, the magneto-electric coupling terms measured here correspond to the terms of second order in electric field within the free energy development: $$-\frac{1}{2}\epsilon_0\epsilon_{ij}E_iE_j -\frac{1}{2}\gamma_{ijk}E_iE_jH_k -\frac{1}{2}\eta_{ijkl}E_iE_jH_kH_l -...$$ The results obtained for the Fe-langasite means that only $\eta_{ijkl}$ contributes to the measured magneto-electric effect. It is worth noting that this coupling term is always present in the free energy development, as it is of second order in both electric and magnetic field, so no particular symmetry conditions determine its existence, unlike for $\gamma_{ijk}$. This quadratic coefficient however varies with the temperature, being maximum close to the N\'eel temperature. This variation reflects the critical behavior of the magneto-electric effect describable in the frame of the fluctuation-dissipation theory\cite{Nugroho}. Other magneto-electric coupling terms are accessible when measuring the magnetic dependence of the electric polarisation (which is the first partial derivative of the free energy with respect to the electric field). Such experiments are underway.

\section{Discussion \label{discussion}}

\subsection{Magnetic properties}

Our investigations of the Fe-langasites magnetic properties have revealed an original magnetic order at low temperature associated to the triangular array of triangles of magnetic Fe$^{3+}$. The magnetic moments are orientated at 120\degre from each other within the triangles and propagate helically perpendicular to the triangular planes. There are at least two energy scales in this system since the intra-triangle super-exchange is expected to be much stronger than the in-plane and out-of-plane inter-triangle interactions which are mediated by two oxygens. The Curie-Weiss temperature mainly reflects this intra-triangle interaction which also gives rise to magnetic fluctuations above the ordering temperature (see figure \ref{BNFS_diff}). The Curie-Weiss temperature is larger than the N\'eel temperature by almost an order of magnitude. The 3 dimensional ordering sets in when the temperature becomes of the order of the super-super exchange paths within the planes and between them. We have previously shown \cite{BNFS_PRL} that this helical arrangement  results from  a twist of the interplane interaction: the strongest interaction among the three that link one Fe$^{3+}$ of a triangle to the three Fe$^{3+}$ of the  triangle of the next plane is diagonal (J3 or J5). If, moreover, the two other interactions are zero, it is easily shown that the $z$-component of the propagation vector would be exactly $\tau$=1/6=0.167, i.e. a rotation of the moments of 60\degre from plane to plane\cite{BNFS_PRL}. The departure of the propagation vector from 1/6 arises when the two other inter-plane interactions are no more negligible. 
\\

In the present study, we have focused on the effect on the magnetic properties of  different non-magnetic cations on sites A, B and D. The most striking result is that all the magnetic characteristics are roughly identical for all compounds even when the substitution produces a large variation of the cell parameters, except for those containing Sb (in place of Nb or Ta). The presence of this cation systematically increases the N\'eel temperature, decreases the Curie-Weiss one, increases the value of the ordered magnetic moment, and increases the $\tau$ value (i. e. shortens the helical modulation period). The origin of this effect could be structural, the Sb$^{5+}$ radius being slightly smaller than the Ta$^{5+}$ or Nb$^{5+}$ ones. However, no systematic trend on exchange path geometry was revealed by the structural investigation performed with powder neutron scattering on chosen compounds. The microscopic mechanism at play should therefore be of electronic origin. It could be related to the larger Pauling electronegativity of Sb$^{5+}$ (2.05) with respect to Nb$^{5+}$ (1.6) and Ta$^{5+}$ (1.5) \cite{Sb}. The departure of this value from the oxygen electronegativity (3.44) gives the degree of ionicity/covalency and polarisation of the cation-oxygen bond involved in the magnetic exchange paths. This should therefore give a more covalent character to the Sb-O bond. The electronic structure of the Sb$^{5+}$ cation is also different from the Nb/Ta ones (different column of the periodic table) with in particular a full outermost 4d shell. 

Remembering that two consecutive Fe$^{3+}$ triangles along the $c$ axis are coupled via oxygen anions forming the octahedral coordination of the D site cation (see figure \ref{Feplan_3}), it is not surprising that a change of the electronic affinities of this cation has consequences onto the magnetic coupling within and between the planes. For instance, the increased covalence of the Sb-O bond seems in turn to decrease the strength of the J1 Fe-O-Fe bond involving the same oxygen ions. The consequence of this is the observed lowering of the Curie-Weiss temperature and smaller spin transfer from the Fe$^{3+}$ to the oxygen (hence a larger ordered moment). 

The influence of the electronic properties of Sb on the N\'eel temperature and on the helix propagation vector is more difficult to track since it implies more complicated exchange paths between two planes.  It is however interesting to note that the angle of rotation of the spin is larger than 60\degre for Sb compounds and smaller  than 60\degre for the Nb/Ta ones. This implies a contribution of the two weaker interplane interactions, whose ratio would be inverted between Sb compounds and Nb/Ta ones.

\subsection{Dielectric properties}

It is of course of interest to address the question of the influence of this increased Sb-O bond covalence on the dielectric properties of the Fe-langasites. The dielectric measurements on all powder samples have not shown significant difference though, with a similar cusp of the dielectric permittivity close to T$_N$, suggesting the same kind of magneto-electric coupling for the different compounds. This comparative study will be carried on with electric polarisation measurements.   

The dielectric properties reported here for the Fe-langasites are strikingly similar to those of  YMnO$_3$. The magnetic structure of this compound presents some resemblance with the Fe-langasite one : a triangular array of in-plane magnetic moments oriented at 120\degre from each other in the ordered phase. The dielectric measurements performed on this multiferroic compound show a comparable anomaly in the thermal variation of the ($a$,$b$) in-plane dielectric permittivity while nothing is visible for the dielectric permittivity measured along the $c$ axis, and a quadratic dependence of the dielectric permittivity close to this anomaly with the magnetic field (note however that in the YMnO$_3$ study, the magnetic field was oriented perpendicular to the electric one). This anisotropic dielectric behaviour of YMnO$_3$ has been described in a phenomenological Landau description of ferroelectric antiferromagnets including a fluctuative part \cite{Nugroho}. Unlike the langasites, YMnO$_3$ is already ferroelectric at temperatures well above the magnetic transition (proper multiferroic). Anyway, the magnetic ordering still induces an additional contribution to the overall electric polarisation. The rise of electric polarisation at the N\'eel temperature has also been claimed very recently in the Fe-langasite by Zhou {\it et. al.} \cite{zhou} from polarization measurements performed along the $c$-axis. This would imply that, at variance with YMnO$_3$, these langasite compounds should be classified as improper multiferroics, the magnetic transition driving a paraelectric-ferroelectric transition. Such behaviour has already been reported in another compound containing an Fe triangular lattice where the electric polarisation is induced at the N\'eel transition in the $c$ direction along the axis of the helix and perpendicular to the plane of the moments oriented at 120\degre from each other \cite{kenzelmann}.

\section{Conclusion \label{conclusion}}

The Fe Langasites are the first compounds of this family to evidence a magnetic ordering whose peculiar arrangement reflects magnetic frustration and the chirality of the structure. This rich family offers several Fe-based compounds with different cationic substitutions among which we have highlighted the singular role of the Sb cation on the magnetic properties. The langasite structure is non-centrosymmetric and a further loss of symmetry elements is achieved in the magnetic structure that leave an overall polar structure authorising  the appearance of a spontaneous electric polarisation. This would then be associated to a structural phase transition as suggested by high resolution x-ray diffraction. Dielectric measurements reveal the presence of magneto-electric coupling through an anomaly at the N\'eel temperature in the thermal variation of the dielectric permittivity and its quadratic dependance with the magnetic field. The Fe-langasite could then be a new example enlarging the class of magnetoelectric/multiferroic triangle-based antiferromagnets.

\begin{acknowledgments}
This work has benefited from the financial support of the ANR contract BLAN06-1 140756. We would like to thank J. Marcus, J. Debray and B. Kundys for their help during the dielectric measurements, and E. Ressouche for his contribution to the single crystal neutron diffraction experiment. 
\end{acknowledgments}


\end{document}